\documentclass[12pt]{article}
\pdfoutput=1
\usepackage{graphicx}
\usepackage{color}
\usepackage{amsmath,slashed, amssymb}
\usepackage{fancyhdr}
\usepackage{hyperref}
\usepackage{url}
\usepackage[utf8]{inputenc} 

\def\beq{\begin{eqnarray}}
\def\eeq{\end{eqnarray}}
\def\bea{\begin{eqnarray*}}
\def\eea{\end{eqnarray*}}

\def\centeron#1#2{{\setbox0=\hbox{#1}\setbox1=\hbox{#2}\ifdim
\wd1>\wd0\kern.5\wd1\kern-.5\wd0\fi
\copy0\kern-.5\wd0\kern-.5\wd1\copy1\ifdim\wd0>\wd1
\kern.5\wd0\kern-.5\wd1\fi}}
\def\ltap{\;\centeron{\raise.35ex\hbox{$<$}}{\lower.65ex\hbox{$\sim$}}\;}
\def\gtap{\;\centeron{\raise.35ex\hbox{$>$}}{\lower.65ex\hbox{$\sim$}}\;}

\def\singleandthirdspaced{\baselineskip=\normalbaselineskip\multiply
    \baselineskip by 130\divide\baselineskip by 100}

\newcommand{\newc}{\newcommand}
\newc{\qbar}{{\overline q}}
\newc{\Kahler}{Kahler }
\newc{\deltaGS}{\delta_{\rm GS}}
\begin{document}
\begin{titlepage}
\begin{flushright}
{\large SCIPP 14/14\\
}
\end{flushright}

\vskip 1.2cm

\begin{center}

{\LARGE\bf Instanton Effects in Three Flavor QCD}

\vskip 1.4cm

{\large Michael Dine$^{(a)}$, Patrick Draper$^{(b)}$, and Guido Festuccia$^{(c)}$}
\\
\vskip 0.4cm
{\it $^{(a)}$Santa Cruz Institute for Particle Physics and
\\ Department of Physics,
     Santa Cruz CA 95064  } \\
\vspace{0.3cm}
{\it
$^{(b)}$Department of Physics, University of California, Santa Barbara, CA 93106
}\\     \vspace{0.3cm}
{\it
$^{(c)}$Niels Bohr International Academy and Discovery Center, Niels Bohr Institute, \\
University of Copenhagen, Blegdamsvej 17, 2100 Copenhagen \O, Denmark
}\\     

\vskip 4pt

\vskip 1.5cm

\begin{abstract}
Recently it was shown that in QCD-like theories with $N_f > N$, 
where $N_f$ is the number of light flavors and $N$ is the number of colors, 
there are correlation functions that vanish in perturbation theory and at short distances receive dominant, calculable contributions from small instantons.  
Here we extend the set of such objects to theories with $N_f = N$, which includes real QCD, and discuss their application as a calibration of lattice computations at small quark mass.  We revisit the related issue of the $u$ quark mass and its additive renormalization by small instantons, and discuss an alternative test of $m_u=0$ on the lattice.
\end{abstract}

\end{center}

\vskip 1.0 cm

\end{titlepage}
\setcounter{footnote}{0} \setcounter{page}{2}
\setcounter{section}{0} \setcounter{subsection}{0}
\setcounter{subsubsection}{0}
\setcounter{figure}{0}

%%%%%%%%%%%%%%%%%%%%%%%%%%%%%%%%%%%%%%%%%%%
%%%%%%%%%%%%%%%%%%%%%%%%%%%%
\singleandthirdspaced

\section{Introduction}

Two particularly interesting questions in QCD are the origin of the $\eta^\prime$ mass and the possibility that a small mass for the $u$ quark might solve the strong CP problem.  Both questions are inherently non-perturbative.  Lattice gauge theory has made enormous strides in the last decade on each:  the $\eta^\prime$ mass is reasonably well reproduced~(see, for example, \cite{Christ:2010dd,Gregory:2011sg,Michael:2013vba}), while the quark masses are known at the $5\%$ level, with simulations bracketing the physical quark masses at lattice spacings of order $(3 ~{\rm GeV})^{-1}$ or smaller (see, for example, the detailed review~\cite{review2}).

The elimination of the possibility that $m_u$ vanishes is particularly important. The other known solutions to the strong CP problem, the Peccei-Quinn and Nelson-Barr mechanisms, both exhibit substantial challenges from the theoretical point of view.  The computations of the light quark masses are quite complex, so it is reassuring that simulations performed by different methods yield similar results~\cite{Bazavov:2009bb,Bazavov:2009fk,Durr:2010vn,Durr:2010aw,McNeile:2010ji,Aoki:2012st,Laiho:2011np}. However, it would be useful to establish independent cross-checks of $m_u>0$ on the lattice, separate from fits of the light quark spectrum. More generally, it would be interesting to have analytic probes of nonperturbative physics that could serve as a calibration of lattice computations sensitive to the chiral anomaly.

At first sight, $m_u=0$ appears inconsistent with results of current algebra, but Georgi and McArthur~\cite{georgimcarthur}, Choi, Kim, and Sze~\cite{choikimsze}, and Kaplan and Manohar~\cite{kaplanmanohar} pointed out reasons why this might be misleading.  In~\cite{georgimcarthur} and~\cite{choikimsze} it was shown that instantons contribute to an effective mass for the $u$ quark at QCD scales, proportional to $m_d m_s$ and an IR-divergent integral over instanton scale sizes.  Ref.~\cite{kaplanmanohar} discussed more generally what can be learned by fitting chiral lagrangians to meson spectra, noting that there are other operators quadratic in masses which transform like the linear terms under the underlying chiral symmetries, and that  these effects are parametrically of order $m_d m_s/\Lambda_{QCD}$, plausibly as large as the naive $m_u$.  Banks, Nir, and Seiberg~\cite{banksnirseiberg} developed these arguments further, clarifying the connection between the chiral lagrangian and the underlying microscopic theory, and discussing the circumstances under which a massless or nearly massless $u$ quark might accidentally emerge from underlying symmetries.

We will study probes of both the low energy constants (LECs) controlling the Kaplan-Manohar operator in the chiral lagrangian and the instanton configurations that contribute to them. In the first part of this work we discuss the dependence of $m_\pi^2$ on $m_s$, which is sensitive to the relevant combination of LECs. We point out that the coefficient of this operator can already be estimated from existing lattice data on the LECs, but the uncertainty is substantial. Interestingly, it {\it is} suppressed in large $N$ \cite{banksnirseiberg,ckn}.  It would be desirable if this parameter could be fit more precisely with well-established systematic and statistical
errors. A small value not only corroborates $m_u>0$, but lends quantitative support to the large $N$  picture as a good description and instantons as less important for describing QCD at low scales.  

In the second part of this work we discuss a more general tool for studying small instantons in QCD and on the lattice. While instantons are suggestive of the origin of the $\eta^\prime$ mass, and can provide a potentially substantial contribution to the $u$ quark mass, their precise role is unclear.  Instanton computations are plagued with infrared divergences, and Witten long ago argued that instantons are not the dominant players in understanding the mass of the $\eta^\prime$~\cite{Witten:1978bc,wittencurrentalgebra,wittencurrentalgebra2}.\footnote{Possible ways in which the instanton and large $N$ viewpoints might be reconciled are discussed in \cite{shuryak,Schafer:2004gy}. For a recent discussion in theories under semiclassical control, see~\cite{Poppitz:2012nz}.} However, there are certain correlation functions in gauge theories that, at short distances, receive dominant, {\it calculable} contributions from instantons.  A limited set of such objects was noted in~\cite{dinefestuccia} for $N>N_f$; here we extend the class of theories to include the phenomenologically relevant case of QCD with three light quarks\footnote{The arguments of Ref.~\cite{dinefestuccia} are self-consistent.  They rely on a set of assumptions explained in that work, and some further elaboration will be provided in this paper.}. These Green's functions provide a set of benchmark, non-perturbative quantities accessible to both analytic calculation and numerical simulation.  At short distances, they should in principle match well between the two. 
At larger separations, they could provide a lattice measure of how the instanton IR divergences are physically cut off.  
The rest of this paper is organized as follows. In Sec.~\ref{uquarkmass} we review the theory of the up quark mass. We repeat (and slightly correct) the instanton calculation of Georgi and McArthur and describe the issue from the perspective of  chiral perturbation theory. In Sec.~\ref{ChPTtest} we then discuss our first test of $m_u=0$ using the linear dependence of the pion mass on the strange quark mass. In Sec.~\ref{greenfunctions} we turn to the family of correlators that probe small instantons in QCD. We establish that small instantons are sensible configurations, not only in theories with $N_f > N$, but also in QCD-like theories, including the case of $N_c = 3 = N_f$, and that the notion of an instanton density is well-defined for small instantons.\footnote{We thank R. Kitano for discussions of his program to extract this quantity by different lattice methods.}
Subsequently we compute the leading semiclassical contribution to a particular Green's function at short distances and analyze subleading corrections, organized with the operator product expansion. We discuss optimal sets of correlators and considerations for lattice simulations before turning to the more speculative question of the role of instantons in quantities where the semiclassical analysis leads to infrared divergences.  The most basic model for such calculations is to introduce a sharp cutoff on the instanton scale size.  We consider the effects of simple cutoffs and note that the finite Green's functions may suggest a lower bound on the cutoff parameter.
 Finally, in Sec.~\ref{howsmall} we consider some of the theoretical issues associated with a small bare $m_u$. We explain that the status of small $m_u$ is similar to that of a high quality Peccei-Quinn symmetry.  It might be an accidental consequence of horizontal symmetries in a theory of flavor, as in~\cite{banksnirseiberg}; alternatively, the low energy
 theory may simply possess apparently anomalous discrete symmetries, a phenomenon
 familiar in string theory\cite{banksdinediscrete}.

%%%%%%%%%%%%%%%%%%%%%%%%%%%%%%%%%%%%%%%%%%%

\section{Review of the Theory and Status of $m_u$}
\label{uquarkmass}

In this section we briefly review the nonperturbative renormalization of the up-quark mass, its relation to the Kaplan-Manohar ambiguity in the chiral Lagrangian, and the status of lattice computations of the light quark spectrum.

\subsection{Nonperturbative Renormalization}

Even if the up-quark mass vanishes somewhere in the ultraviolet, symmetries permit a nonperturbative additive renormalization of the form
\begin{align}
\partial_t m_u=\gamma m_u + C(g)a m_d^* m_s^*\;,
\label{eq:muRGE}
\end{align}
where $\gamma$ is the perturbative anomalous dimension and $a$ is the Wilsonian length cutoff.
Small instanton contributions to second term in this RGE were first discussed in Refs.~\cite{georgimcarthur,choikimsze}. The instanton computations suffer an IR divergence in the integral over instantons sizes.  As an estimate, we can compute the contribution to $m_u$ from instantons with size less than a sharp cutoff, $\rho <\rho_0$.  Of course, this computation does not capture the complete set of corrections to the Wilsonian $m_u$, 
and may not be the dominant contribution,
but it can give us a sense of the order of magnitude of QCD corrections.  

We find, for the correction between the charm threshold and $\rho_0$,
\begin{align*}
m_u[\rho_0] = 1.15 {8 \pi^6\over 3}\tilde{\Lambda}^9 \frac{ m_s(\mu) m_d(\mu)}{\alpha(\mu)^{22/9}} \int_{m_c^{-1}}^{\rho_0} d\rho \rho^9 \left(\frac{\alpha(\mu)}{\alpha(\rho)}\right)^{22/9} 
\left ({\alpha(\rho) \over \alpha(\mu)} \right )^{8/9} \left ({\alpha(\rho_0) \over \alpha(\rho)} \right )^{4/9} +m_u[m_c^{-1}]~.
\end{align*}
In this expression we include factors of $\alpha^{\gamma/\beta}$ generated by resumming higher-loop perturbative corrections at leading log (solving Eq.~(\ref{eq:muRGE})). 
This expression differs from that in~\cite{georgimcarthur}, which only included some of the higher-order perturbative corrections, resulting in a numerically rather different effect as a function of $\rho_0$.

\begin{figure}
\begin{center}
\includegraphics[width=10cm]{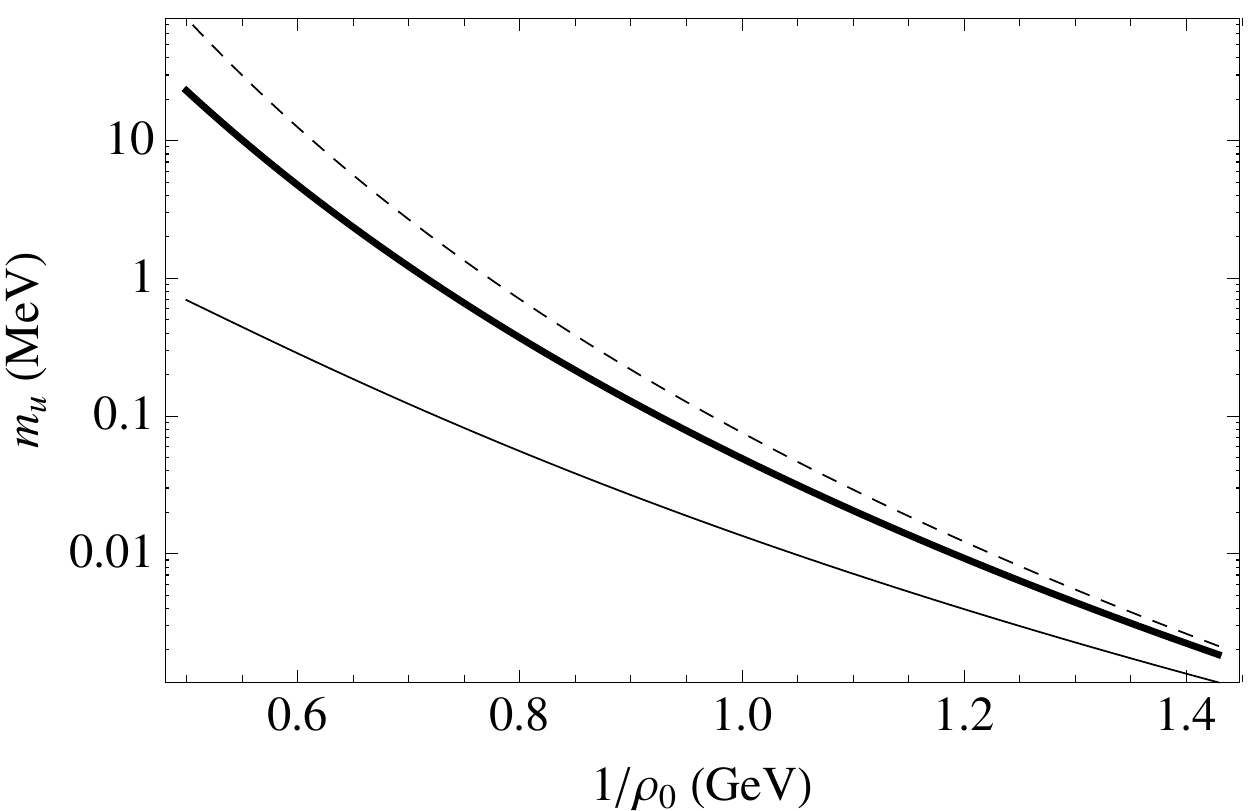}
\caption{The effective $m_u$ generated by small instantons as a function of a hard IR cutoff on the instanton size. Thick line: RG-improved result. Thin line: Georgi-McArthur approximation (partial RG-improvement). Dashed line: one loop result. In all cases we take $m_s=93$ MeV, $m_d=4$ MeV, set the renormalization scale to 2 GeV, and take the limit of a large UV cutoff, ignoring small corrections from heavy-quark thresholds.}
\end{center}
\end{figure}

The additive contribution to $m_u$ is shown in Fig. 1 as a function of $\rho_0$.  If
$\rho_0$ is as small as $0.8$ GeV$^{-1}$, roughly the charm threshold,
the contribution to $m_u$ from smaller instantons is less than a hundredth of an MeV; $m_u = 2$ MeV corresponds to $\rho_0 = 1.5$ GeV$^{-1}$. 

Although our computation improves on that of~\cite{georgimcarthur} for ultraviolet $\rho_0$, due to the strong IR sensitivity, it is still not possible to draw any sharp conclusion about the full nonperturbative contribution to the running $m_u$. We can only conclude, as~\cite{georgimcarthur} did\footnote{Note that in~\cite{georgimcarthur} the limit $\rho_0\rightarrow\Lambda^{-1}$ was taken.}, that it is plausible {\it a priori} that instantons and other nonperturbative effects could contribute $\mathcal{O}(1)$ MeV to $m_u$. 

\subsection{Chiral Perturbation Theory}
Although one might hope to test the nonperturbative renormalization in Eq.~(\ref{eq:muRGE}) with meson phenomenology and second-order chiral perturbation theory, Kaplan and Manohar (KM) pointed out a significant obstacle~\cite{kaplanmanohar}, exhibiting an ambiguity in the parametrization of the chiral lagrangian.  The leading-order term in the lagrangian is:
\beq
{\cal L}_2 = {F^2 \over 4} {\rm Tr} \left (\partial_\mu U^\dagger \partial^\mu U \right ) + {F^2 \over 4} {\rm Tr} \left ( \chi^\dagger U \right ) + {\rm c.c.}
\eeq
where $\chi$ and $U$ are given by
\beq
\chi=2MB_0\;,~~~U = e^{i{ \lambda^A \pi^A \over F}} ~,
\eeq
$M$ is the quark mass matrix, and $B_0$ is proportional to the magnitude of the chiral condensate. Second-order terms are parametrized by the Gasser-Leutweyler (GL) parameters $L_{1-8}$~\cite{gl}. In brief, the KM ambiguity is the statement that there is a particular combination of operators (with the quantum numbers of, and receiving contributions from, small instantons) that has the potential to mimic the effects of a non-zero bare $u$ quark mass.  Following~\cite{banksnirseiberg}, the operator can be written as
\beq
\mathcal{L}\supset r_1 \left ( {\rm Tr} (\chi^\dagger U \chi^\dagger U) - {\rm Tr}  (\chi^\dagger U)^2 \right ),
\eeq
where in terms of the GL parameters,
\beq
r_1 = \frac{1}{2}\left(L_8-L_6-L_7\right)\;.
\eeq
By a redefinition of $\chi$, $r_1$ can be eliminated, providing an effective contribution to $m_u$ of order $m_dm_s$.    Alternatively, having fixed the ambiguity by requiring -- for instance -- that $M$ is proportional to the UV quark mass matrix, a large value of $r_1$ and a small value of the bare $m_u$ would be compatible with the observed pseudoscalar meson masses, whereas the orthogonal combinations of GL parameters are fixed by the spectrum. An $r_1$ of order $10^{-3}$ would be sufficient if $m_u=0$. A non-zero $r_1$, with orthogonal combinations of $L$'s comparatively smaller, corresponds to:
\beq
-2 L_6 \approx -2 L_7 \approx L_8 \approx r_1.
\eeq

\subsection{Lattice QCD}

The only tool we have to reliably determine the light quark masses -- and in particular whether the $u$ quark mass is nearly zero in the UV -- is lattice gauge theory.  Light
quarks are perhaps the biggest challenge for the lattice,
but over the past decade, lattice computations have yielded remarkably precise values for their masses.  The FLAG  review~\cite{review2} summarizes the results from several collaborations, and they are generally in good agreement, giving values for $m_u$ ($m_d$) of order $2$ MeV ($4.5$ MeV) at a scale of 2 GeV, with systematic and statistical errors around $5\%$. From these results, $m_u$ deviates from zero with high statistical significance. In the remainder of this work, we discuss two methods of cross-checking of $m_u>0$, orthogonal to direct fits of the light quark specturm.

\section{Testing $m_u=0$ with Low Energy Constants}
\label{ChPTtest}
As we will discuss in this section, testing the $m_u=0$ hypothesis does not require obtaining precise values for $m_u$ and $m_d$.
This question can be addressed with meson spectra in lattice simulations away from the physical point, and in particular their variation with the quark masses. 

The critical point is that the KM transformation is not a symmetry of QCD, and the lattice can resolve it by measuring some quantity sensitive to $r_1$. For example, consider corrections to the average pion mass proportional to $m_s$, 
\beq
m_\pi^2 = \beta_1 (m_u+m_d) + \beta_2 m_s (m_u+m_d)+\mathcal{O}(m_{u,d}^2)\;.
\eeq
Lattice calculations are often done with $m_{u}(a) = m_{d}(a) \equiv \hat m$. The parameters $\beta_1$ and $\beta_2$ can be extracted on the lattice by varying $\hat m$ and $m_{s}(a)$  independently; e.g.,
\beq
{\beta_2 \over \beta_1} \approx \frac{m_{\pi_1}^2-m_{\pi_2}^2}{m_{\pi_2}^2m_{s_1}-m_{\pi_1}^2m_{s_2}}\;.
\label{beta2overbeta1}
\eeq
if simulations are done at two values of $m_s$.\footnote{A related measurement was discussed in~\cite{ckn} as a method to fix the ambiguity.} Taking the simplified limit where GL parameters orthogonal to $r_1$ are negligible, the $\beta_i$ reduce to
\beq
\beta_1 =  B_0\;,~~~\beta_2 = -16 {r_1 B_0^2 \over F^2}\;.
\eeq
In this limit, if $m_u$ vanishes, then the combination of $m_\pi^2$ and the kaon masses can be used to formulate the constraint
\beq
{\beta_2\over \beta_1} \approx \frac{1}{m_s}\frac{m_\pi^2-(m_{K^0}^2-m_{K^+}^2)_{QCD}}{m_\pi^2+(m_{K^0}^2-m_{K^+}^2)_{QCD}}\;,
\label{betaconstraint}
\eeq
where the subscript indicates that only the QCD contribution to the kaon splitting is used. Numerically, this constraint gives
\beq
{\beta_2 \over \beta_1} \approx 5~{\rm GeV}^{-1}.
\eeq
Corrections to Eq.~(\ref{betaconstraint}) from keeping the other GL parameters are easy to include.  More precisely,
\beq
{\beta_2 \over \beta_1} = 16{B_0 \over F^2} (2L_6-L_4)~,
\eeq
and the same combination of the $L_i$ as well as the combination $2L_8-L_5$ appear again in the kaon masses, so we can write a more general formula relating $\beta_2/\beta_1$ to  $m_\pi^2$, $m_K^2$, $m_s$, and $B_0$. This relation 
increases the required $\beta_2 \over \beta_1$ to $\mathcal{O}(10)$ GeV$^{-1}$. Alternatively, if the quadratic dependence of $m_K^2$ on $m_s$ is measured on the lattice, the more general constraint can be written in the form
\beq
{\beta_2\over \beta_1} \approx \frac{1}{m_s}\frac{m_\pi^2-(m_{K^0}^2-m_{K^+}^2)_{QCD}}{m_\pi^2+(m_{K^0}^2-m_{K^+}^2)_{QCD}}+\frac{1}{B_0}\frac{m_\pi^2 (\partial^2m_K^2/\partial m_s^2)}{m_\pi^2+(m_{K^0}^2-m_{K^+}^2)_{QCD}}\;.
\eeq
From the results quoted in~\cite{review2} for $B_0$, $m_s$, and $2L_6-L_4$, we can estimate
\beq
{\beta_2 \over \beta_1} \simeq (1\pm 1)~{\rm GeV}^{-1}.
\eeq
Although the error bars are large (and here only crudely estimated), the ratio is too small to account for the effects of the $u$ quark mass. But $\beta_2/\beta_1$ is a fundamental prediction of QCD and it would be interesting to see a dedicated study with increased precision. It would provide another demonstration of $m_u\neq0$, as well as a probe of the contribution of small instantons to the chiral lagrangian.

\section{Instantons and Nonperturbative Green's Functions}
\label{greenfunctions}

We turn now to a more general test of nonperturbative physics on the lattice, which is sensitive to the same short-distance gauge field configurations that renormalize $m_u$ in the UV.

In~\cite{dinefestuccia} it was observed that in gauge theories with $N_f >N$ massless flavors, certain Green's functions vanish in perturbation theory, and at short distances receive a contribution from an instanton that is infrared-finite and calculable in a systematic expansion in $\alpha(x)$.  For the case of $N=2, ~N_f=3$, for example, one such Green's function behaves as
\beq
\langle \bar u d \bar d s(x) ~\bar s u(0) \rangle \sim \Lambda^{16/3} x^{-11/3} + ~{\rm nonsingular}.
\label{nfgreaternc}
\eeq
The one-instanton computation generating the singular term in Eq.~(\ref{nfgreaternc}) is both infrared and ultraviolet finite, and perturbative corrections can be computed. It was argued in~\cite{dinefestuccia} that IR-divergent corrections from the instanton ensemble do not contribute to the leading singular behavior. 

The operator product expansion helps clarify the UV and IR structure. The OPE for the operator in Eq.~(\ref{nfgreaternc}) has the form
\begin{align}
\bar u d \bar d s(x)~ \bar s u(0) \sim \big(c \Lambda^{16/3} x^{-11/3}+{\rm nonsingular} \big){\rm \bf I}~+\big(1 + {\cal O}(\alpha)\big) {\rm :}\bar u d \bar d s \bar s u(0){\rm :} + \dots
\label{OPE1}
\end{align}
The coefficient of the unit operator is the sum of a {\it singular} contribution, which can be computed systematically in perturbation theory about a single instanton, and {\it nonsingular}, incalculable corrections generated by interactions with the full instanton ensemble~\cite{dinefestuccia}.  The coefficient of the six-fermion operator is nonsingular.  In the one-instanton background, its matrix element is UV divergent, so the operator must have a subtraction applied as denoted by the normal-ordering in Eq.~(\ref{OPE1}).\footnote{Infrared finiteness of certain \emph{matrix elements} like this one in a one-instanton background has been discussed in various works~\cite{Shuryak:1987an,Velkovsky:1997fe,Zhitnitsky:2013wfa}, and correctly noted to acquire incalculable contributions from the instanton ensemble.}  With connected dilute gas corrections, the matrix element acquires IR divergences and is not calculable analytically. In principle it can be computed numerically, for example, on the lattice.  In any case, in each order of the perturbation expansion, we expect that the \emph{most singular term} in~(\ref{nfgreaternc}) is calculable.

The demonstration given in~\cite{dinefestuccia} that the (one-instanton, instanton-ensemble) contributions to the correlation functions factorize as above into (most singular, less singular) terms falls short of a rigorous proof. It is an assumption of this work that this factorization holds.

We can generalize to other operators, replacing, for example, $\bar u d(x)$ by $\bar u\sigma_{\mu \nu} F^{\mu \nu} d(x) $.  The unit operator coefficient is now more singular by two more powers of $x$.  Similarly, the six fermion operator also appears, now with a singular coefficient proportional to $1/x^2$ and powers of $(\alpha/\pi)$.  But again the most singular term is the unit operator and it remains calculable.

It is interesting to consider whether other theories, and in particular $N=N_f=3$, possess Green's functions with similar properties.

\subsection{Green's functions in pure $SU(N)$}
\label{calculable}

Based on the finite Green's functions described above, one might hope to find similar objects in other theories.  Consider, for example,
pure ($N_f =0$) $SU(2)$ gauge theory.  In perturbation theory, the Green's function
\beq
G(x) = \langle F^2(x) ~F \tilde F(0) \rangle
\eeq
vanishes as a result of CP invariance.  In an instanton background, with a nonzero vacuum angle $\theta$, $G$ is proportional to $\sin \theta$:
\beq
G(x) = c \Lambda^{22/3} x^{-2/3}\sin \theta.
\eeq
The leading instanton contribution is infrared finite and mildly singular for small $x$.  However, higher-order corrections, although suppressed by $\alpha$, are more singular at short distances.   In the OPE description, the operator $F\tilde{F}$ appears:
\beq
 F^2(x)~ F \tilde F(0) \sim c \Lambda^{22/3} x^{-2/3}\sin \theta\; {\rm \bf I} + k x^{-4} F \tilde F(0) + \dots
\eeq
Although $k$ is $\mathcal{O}(\alpha/\pi)$, at sufficiently short distances, the $F\tilde{F}$ contribution dominates over that of the unit operator.  Moreover, the expectation value of $F \tilde F$ is incalculable (unless $\theta=0$, in which case it vanishes along with the rest of $G$). Its leading instanton contribution diverges as the $10/3$ power of any would-be infrared cutoff.

In a lattice computation (capable of measuring $\theta$-dependent effects), the unit operator might be isolated by working at moderate (not extremely small) $x$  and subtracting $k\alpha x^{-4}$ times a lattice-measured value of $\langle F \tilde F \rangle$.  But at the very least the procedure would be extremely challenging.  The OPE structure in this example is general among pure gauge theories, as well as theories with $N_f<N$. At best, the only computable quantities in these theories are described by subleading terms in an operator product expansion.  

\subsection{$N_f = N$}

As discussed in \cite{dinefestuccia}, $N_f = N$ is a borderline case for fermionic correlators analogous to Eq.~(\ref{nfgreaternc}). They are not strictly calculable in a 1-instanton background, possessing logarithmic IR divergences that correspond in the OPE to matrix elements of the multiquark local operators. 
However, these theories are particularly interesting both because of the relevance to nature for $N_f=N=3$ and because there is a wealth of lattice data. 

Correlators of operators with field strength insertions provide a more effective probe.  In the $N_f=N=3$ case, for example, consider the Green's function
\beq
G(x) = \langle \bar u \sigma_{\mu \nu} F^{\mu \nu} d(x) ~\bar d  \sigma_{\rho \sigma} F^{\rho \sigma} s \bar s u(0) \rangle.
\label{n3nf3}
\eeq
$G$ is finite in the 1-instanton background at leading order in $\alpha$. Correspondingly, the OPE includes the unit operator with a $\Lambda^9 x^{-4}$ singularity. 
$G$ acquires an infrared divergence when the gauge fields are allowed to fluctuate due to the contraction of $F(x)F(0)$ in the correlation function.  The OPE takes the form:
\begin{align}
\label{ex3f}
\bar u \sigma_{\mu \nu} F^{\mu \nu} d(x) ~\bar d  \sigma_{\rho \sigma} F^{\rho \sigma} s \bar s u(0)\sim c\Lambda^9 x^{-4}\big(1+k\log(x\mu)\big){\rm \bf I}+k x^{-4} {\rm :}\bar u u \bar d d \bar s s(0){\rm :}+\dots
\end{align}
where $k$ is $\mathcal{O}(\alpha/\pi)$. Here the six quark operator is schematic and stands for a family of similar operators with different spin contractions. The $\log(x\mu)$ term in the coefficient of the unit operator includes nonperturbative operator mixing, generated by the logarithmic UV divergence of $\langle \bar u u \bar d d \bar s s(0)\rangle$ in the instanton background. 

As mentioned above the six quark matrix elements are also IR log-divergent and incalculable. However, their coefficients are not more singular than that of the unit operator, and furthermore $k$ is $(\alpha/\pi)$-suppressed. Therefore, the calculable term is much easier to isolate. As a first approximation, the incalculable matrix elements might simply be ignored: at scales of order $x\sim m_\tau^{-1}$, for example, $\alpha/\pi$ is a 10\% effect. More accurately, in a lattice computation they could in principle be measured and subtracted from $G$.

We chose the form of $G$ in Eq.~(\ref{n3nf3}) because it exhibits simply how such Green's functions may be decomposed into calculable and incalculable terms. With slightly different choices of $G$, the incalculable matrix elements can be pushed off to even higher orders in perturbation theory. For example:
\begin{align}
&\langle \bar u \sigma_{\mu \nu} F^{\mu \nu} d(x)~ \bar d   s(y) ~\bar s u(0)\rangle\;,\nonumber\\
&\langle \bar u \sigma_{\mu \nu} F^{\mu \nu} d(x)~ \bar d  \sigma_{\rho \sigma} {\tilde F}^{\rho \sigma} s(y) ~\bar s u(0)\rangle\;,\nonumber\\
&\langle \bar u \sigma_{\mu \nu} F^{\mu \nu} d(x) ~\bar d  \sigma_{\rho \sigma} F^{\rho \sigma} s(y) ~\bar s  \sigma_{\lambda \pi} F^{\lambda \pi} u(0)\rangle\;,\nonumber\\
&\langle uds F^{\mu \nu} (x) ~\bar u \bar d \bar s  {\tilde F}_{\mu\nu} (0)\rangle\;,
\label{bettercorrelators}
\end{align}
all have OPEs with the six-fermion operators appearing at $\mathcal{O}(\alpha/\pi)^2$. Note that in Eq.~(\ref{bettercorrelators}) we have also separated the operators in a way that prevents disconnected contributions to the correlators.

Returning to the Green's function in Eq.~(\ref{n3nf3}) for illustration, we can evaluate the contribution in the instanton background at leading order in $\alpha$ (the coefficient $c$):
\begin{align}
G(x)= -144 \left({2 \over \pi^2}\right)^3 \int  {d^4 x_0d\rho \over \rho^5} C(g)  (\Lambda \rho)^9
\left({\rho \over (x-x_0)^2 + \rho^2}\right)^5  \left( {\rho \over x_0^2 + \rho^2 }\right)^8 ~.
\end{align}
Combining denominators with Feynman parameters, we obtain
 \beq
G(x)= -{2\over 7} \left({2 \over \pi^2}\right)^2 C(g) \; \Lambda^9 \vert x \vert^{-4}~.
\label{finiteG}
\eeq
The functional determinant calculation fixes the coefficient $C(g)$~\cite{thooftdeterminant}. For the case $N_f = N= 3$ (in the $\overline{MS}$ scheme), we obtain
\beq
C(g) \Lambda^9 \rho^9 = \Lambda^9 \rho^9 \alpha^{-6} 2 \pi^4 e^{4 \times 0.146 - 0.44}~.
\label{eq:Cg}
\eeq
$\Lambda$ is the one loop renormalization group invariant scale, and the numerical factor from the exponent is ${1.15}$. Inserting Eq.~(\ref{eq:Cg}) into Eq.~(\ref{finiteG}) yields the complete singular contribution to the Green's function of Eq.~(\ref{n3nf3}) at leading order in $\alpha$.

 At one loop, the scale of the $\alpha^{-6}$ factor is arbitrary.  
We can define a two loop RG-invariant scale:
\beq
\tilde{\Lambda}^9  = \mu^9 e^{-{2 \pi \over \alpha(\mu)}} \alpha(\mu)^{-{32 \over 9}}\;.
\eeq
Setting $\mu = m_\tau$ and $\alpha(m_\tau) = 0.32$ yields
$\tilde{\Lambda} = 0.333 ~{\rm GeV}$. Then we can write the determinant as
\beq
C(g)\Lambda^9\equiv\tilde C(g) \tilde{\Lambda}^9=1.15 \times (2\pi^4) \times \alpha^{-22/9}\times \tilde{\Lambda}^9 .
\eeq
The scale of the residual coupling factor can be generally be determined in each Green's function from renormalization group considerations.
For instance, in Eq.~(\ref{finiteG}), radiative corrections must remove most of the renormalization scale dependence introduced by the factor $\alpha^{-22/9}$, running $\alpha$ to a scale of order $x$ and leaving only $\mu$-dependence generated by the anomalous dimension of $G$ (a matrix in the presence of operator mixing). However, the variation of the one loop prediction with $\mu$ is logarithmic, and higher-precision calculation will only valuable if it is shown that the lattice can compute the correlators with $\mathcal{O}(50\%)$ precision.

\subsection{Finite Green's Functions and Lattice Tests}
\label{latticetests}

The finite correlation functions provide a potentially interesting arena for lattice gauge computations.  First, they are inherently non-perturbative and test an interesting aspect of lattice simulations.  Second, they are sensitive to phenomena that are important to understanding hadronic physics in the chiral limit.  For example, a computation of these Green's functions on the lattice could be used to constrain a variety of models for possible infrared cutoffs on the instanton size.  In particular, consider a hard cutoff, $\rho_0$.  If, at scales of order $\vert x \vert = 1.5 ~{\rm GeV}^{-1}$, the semiclassical expansion for $G(x)$ is at least as good as perturbation theory (i.e. $G(x)$ is equal to the semiclassical value to order $\alpha \over \pi$, or about $90\%$), then we would obtain a rather weak requirement on the infrared cutoff,
\beq
\rho_0 \gtrsim m_c^{-1}\;. 
\label{limit}
\eeq
If the instantons cut off more softly, the same criterion yields more stringent constraints.  For example, with an exponential cutoff, $e^{-\rho/\rho_0}$, one finds $\rho_0 \gtrsim {\Lambda} ^{-1}$.

However, the effects of UV instantons are inherently small, and one can ask whether they are observable.

Among the challenges to measuring such instanton dominated Green's functions are the effects  of
finite quark masses, which yield perturbative contributions.   On the other hand, current simulations achieve quite small masses, less than $10$ MeV for light quarks and $100$ MeV for the strange quark, and quite small lattice spacings, $a^{-1} \simeq 4~{\rm GeV}$ in some cases.   In any simulation it would be important to choose the Green's function carefully so as to avoid disconnected parts, as in the correlators of Eq.~(\ref{bettercorrelators}).

Take the case $\langle \bar u \sigma_{\mu \nu}F^{\mu \nu} d(x)~\bar d \sigma_{\rho \pi}F^{\rho \pi} s(y)~\bar s u(0) \rangle.$ The leading perturbative contribution behaves as 
\beq
\frac{m_d m_s m_u} {(4\pi^2)^4x^{10}}\;,
\eeq
which can be compared with the non-perturbative contribution appearing at zeroth order in quark masses,
\beq
C(g) \Lambda^9 x^{-4}\;.
\eeq
The latter term is already dominant for $x \gtrsim{ (30\Lambda)^{-1}}$, for quark masses in the range above.  

Typically, for a fixed gauge configuration in a simulation ensemble, the correlator receives a contribution of order the quark masses.  Occasionally,  a configuration will contribute a much larger value (by a factor of order $1/(x^3 m_u m_d m_s)$).  For very small quark masses, the probability to find the latter configurations in an ensemble could be suppressed by the fermion determinant. For very large ensembles, this effect cancels out in Green's functions (due to the $1/m$ behavior of the quasi-zero mode contribution to the fermion propagator in the instanton background), but for smaller ensembles, it may be more convenient to remove the suppression of the probability by hand, and restore it later as a weight for the contribution of each configuration.

%%%%%%%%%%%%%%%%%%%%%%%%%%%%%%%%%%%%%%%%%%%

\section{Theoretical Issues Associated with $m_u=0$}
\label{howsmall}

We conclude with a short discussion of the $m_u=0$ proposal from a theoretical perspective, commenting briefly on two issues: whether $m_u=0$ is well-defined, and whether it is well-motivated.

First, a question raised in some of the lattice literature is whether $m_u = 0$ has an unambiguous meaning~\cite{creutz} (for a recent discussion, see the lecture notes of Sharpe\footnote{\url{http://faculty.washington.edu/srsharpe/brazil13/sharpe_brazil4.pdf}}). A concise counterargument to~\cite{creutz} was given in~\cite{Srednicki:2005wc}; here we add a few additional comments. In part, one's view of this question is shaped by what one may view as possible in lattice computations in practice and in principle.  With the ability to compute with arbitrarily small lattice spacing and small quark mass, the answer is clearly yes, at least in principle, as can be seen from the following.  Suppose the lattice spacing is extremely small, say $a^{-1} = {\rm TeV}$.  Instanton contributions to masses from smaller scales are completely negligible, as can be seen from Fig.~1.  On this lattice, calculate a correlator such as 
that of two currents, 
$\langle \bar{u}\gamma^\mu u(x) \bar{u}\gamma_\mu u(0)\rangle$,
at distances a few times $a$.  
The lattice must reproduce the massless perturbative result, with corrections behaving as $m_u^2 x^2,~m_u^2 a^2$~(or possibly $m_u a$).  
The question of whether the corrections vanish as $m_u \rightarrow 0$ is well-posed.  Of course in practice there may be numerical issues in achieving the required accuracy; this would seem to be a question, however, of the size and nature of systematic errors.

A more interesting question is precisely how small $m_u$ has to be to solve the strong CP problem, and what would be required to establish a suitable bound.  The chiral perturbation theory formula for $d_n$, the neutron electric dipole moment \cite{Crewther:1979pi}, involves ratios of current quark masses.  But what masses are these?  

To answer this question, we first sharpen what is meant by ``instanton" contributions to the quark masses by considering the question of $\theta$-dependence.  If at scale $a$, one presents the lagrangian with $\theta$ appearing in front of $F \tilde F$ only, the contribution to $m_u$ from instantons at scales larger than $a$ is proportional to $e^{i \theta}$.   This piece does not contribute to $d_n$. We lump together all contributions of this type (e.g. from dilute gas corrections to the instanton, but more generally from unspecified non-perturbative sources) to define the ``instanton" contribution.

The masses appearing in the usual expression for $d_n$ clearly do not include the instanton contribution, and if one chooses too small an energy scale, separating these out is problematic.  The simplest procedure (conceptually) is to choose the scale high enough that the contribution to $m_q$ from instantons at shorter distances can be neglected.  In this situation, we require
\beq
{m_u \over m_d} < 10^{-10}.
\eeq

With the sort of lattice spacings achievable at present, however, instanton contributions to $m_u$ (again the contributions from integrating out instantons at scales smaller than $a$) are much larger than $10^{-10}~m_d$.   Spacings of roughly $a^{-1} \approx 10$ GeV or smaller are required.

So there are two ingredients to establishing that $m_u$ is small enough to solve the strong CP problem.  The lattice spacing must be small enough that instanton contributions to $m_u$ from shorter scales are much smaller than $10^{-10}~m_d$, and the value of the mass at $a$ must be smaller that $10^{-10} ~m_d$.

Of course, lattices so small and analyses so precise would be very difficult to achieve, and the analysis would require treatment of the chiral lagrangian to very high order.  At best, one could hope for qualitative evidence that the $u$ quark mass vanishes, but it would be unrealistic to {\it prove} that a small $u$ quark mass was responsible for the solution of the strong CP problem.

Separately, one can ask what is required of an underlying theory to obtain such a small $m_u$.  A similar question arises for the axion solution to the strong CP problem:  how might one obtain a Peccei-Quinn symmetry of adequate {\it quality}~\cite{dineaxions} to solve the strong CP problem.  It is clearly interesting to compare these questions to establish, at a purely theoretical level, whether one or the other solution is more plausible.

One formulation of the problem of a massless $u$ quark was provided in~\cite{banksnirseiberg}.  The authors considered possible non-anomalous symmetries spontaneously broken by an order parameter $S$.  Assuming the symmetry to be discrete, the $u$ quark mass (Yukawa coupling) should be suppressed relative to other quark masses by powers of $S$.  If $S$ is of order, say, CKM angles, suppression by many powers of $S$ is needed, and thus a large or complicated discrete symmetry.  As is well known, the situation for axions is similar.  If the Peccei-Quinn symmetry is broken by an order parameter $\phi$, then if, say, $\phi \sim 10^{11}$ GeV, one needs to suppress operators such as $\phi^{N+4} \over M_p^N$, for quite large $N$ ($11$ or $12$).  If implemented with discrete symmetries, again, large symmetries are required.

Neither of these solutions seems terribly plausible.  A more compelling framework is provided by string theory, where Peccei-Quinn symmetries controlled by small quantities like $e^{-{8 \pi^2 \over g^2}}$ are familiar~\cite{wittenaxion}.  The problem becomes explaining the appearance of the small exponentials, but these are at least possibly required by other considerations.  Such small exponentials can also explain a small $m_u$; anomalous discrete are indeed familiar in string theory~\cite{banksdinediscrete}.  So, in this framework, both solutions of the strong CP problem have a level of plausibility, tied to the existence (or not) of small exponential factors.  We might, tentatively, argue that $m_u=0$ is slightly less plausible.  The axion solution simply requires an extremely small exponential; small $m_u$ requires a small exponential {\it and} an approximate discrete symmetry.  In addition, states with light axions might have the additional virtue 
of possessing a dark matter candidate.

%%%%%%%%%%%%%%%%%%%%%%%%%%%%%%%%%%%%%%%%%%%

\section{Conclusions}
\label{conclusions}

In the first part of this paper we have argued that it would be interesting to have detailed lattice fits to the parameter $\beta_2/\beta_1$ controlling the $m_s$ dependence of $m_\pi^2$. Establishing a bound significantly smaller than $5 ~{\rm GeV}^{-1}$ provides an exclusion of the massless $u$ quark hypothesis that is independent of the direct fits of the light quark spectrum.  We have provided a rough estimate based on published data, and it appears to be five times too small to allow for a massless $u$ quark.   But this is a fundamental QCD parameter, and a dedicated analysis by the different collaborations would be highly desirable.   This quantity can be reliably obtained working with $\tilde m = m_u = m_d$, and quark masses significantly larger than their values in nature.

In the second part of our work we have discussed analytic probes of nonperturbative physics on the lattice more generally. 
We have seen that there are a set of correlation functions in QCD for which, at short distances, instantons provide reliable results, and we have argued that evaluation of such correlators on the lattice would provide a useful calibration.  Such computations may eventually be achievable, given the small lattice spacings and quark masses currently accessible.  Alternatively, the existence of these correlators can be viewed as a demonstration that small instantons are physically meaningful, and in principle they provide a way to extract the instanton density and the IR cutoff on $\rho$ from lattice computations.

\vspace{1cm}

\noindent
{\bf Acknowledgements:}  This work was supported by the U.S. Department of Energy grant number DE-FG02-04ER41286.  G.F. is supported by a Mobilex grant from the Danish Council for Independent
Research and FP7 Marie Curie Actions - COFUND (grant id: DFF - 1325-00061). We thank Tom Banks, Claude Bernard, and Nathan Seiberg for very helpful conversations, We also thank Stephen Sharpe for conversations and comments on the manuscript, and Ryuichiro Kitano for explaining to us his program to measure instanton related effects on the lattice.

\appendix
\section{Connections with the $U(1)$ problem}
\label{uoneproblem}

Witten has argued that - given the qualitative successes of large $N$ ideas in understanding QCD - instantons are unlikely to provide a useful understanding of the $\eta^\prime$ mass and other phenomena. Still, we have seen that small instantons are meaningful, and it is interesting to consider a model where both the $\eta^\prime$ mass and $m_u$ receive contributions from small instantons, suppressed in the infrared by a rigid cutoff on $\rho$. Correlating the two masses, of course, requires that the cutoff is the same in both cases.

If the cutoff is not too large, then the $\eta^\prime$ is the pseudo-Goldstone boson of a $U(1)$ symmetry, which gains mass as a consequence of the anomaly (this, of course, has parallels with the large $N$ treatment).  The Goldstone bosons are then described by a unitary matrix,
$$U = e^{{i \over f_\pi}(\pi^a \sigma^a + \eta^\prime)}.$$
 The effective action for $U$ contains terms of the form:
\beq
{\cal L} = f_{\pi}^2\left( \mu{\rm Tr}( MU) + a~{\rm det} (U)\right)~.
\eeq
The latter term receives contributions from small instantons.  The use of instantons here is not in the spirit of large $N$ (as stressed in \cite{wittencurrentalgebra,wittencurrentalgebra2});  we are seeking, at most, a crude connection between the $m_u$ and $\eta^\prime$, and any detailed statement must be taken with a grain of salt.

\begin{figure}
\begin{center}
\includegraphics[width=10cm]{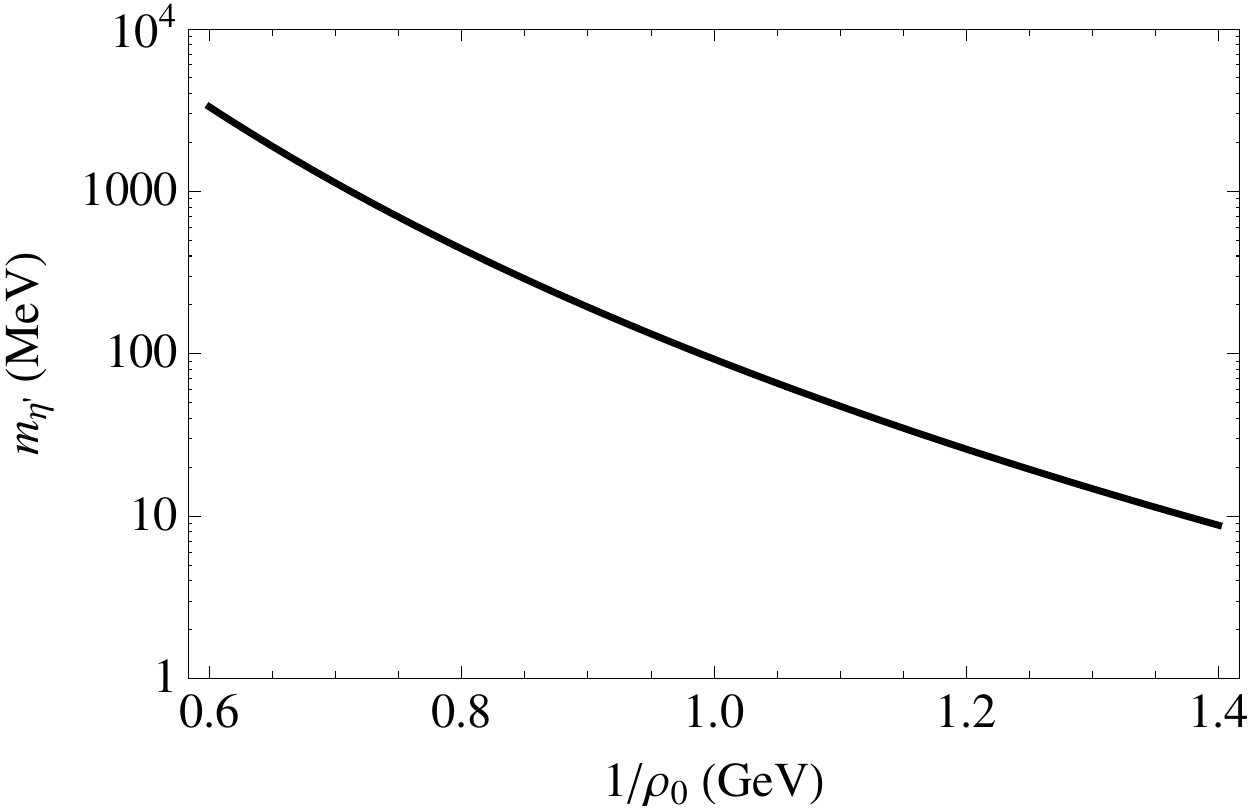}
\caption{Contributions to $m_{\eta^\prime}$ from small instantons as a function of a hard IR cutoff. In contrast to Fig.~1, the estimate here is taken only at 1 loop and will have small logarithmic corrections.}
\end{center}
\end{figure}

From the perspective of the instanton computation, the second term is the 't Hooft interaction, proportional to
$\bar u u \bar d d \bar s s$.  To connect this with the operator $U$, we take $\langle \bar u u \rangle = (250 ~{\rm MeV})^3$ and replace the six quark operators by a simple product.
The contribution from small instantons is very cutoff dependent.  But, except for very large cutoff $\rho_c$, it will not give an appreciable contribution to the $\eta^\prime$ mass, as seen in Fig.~2.  It may be hard to make sense of this calculation for any cutoff below $m_{\eta^\prime}\sim1$ GeV, and the cutoff required to generate the full ${\eta^\prime}$ mass is approximately $1/(0.7)$ GeV.  The same cutoff applied to the $u$ quark mass computation would lead to a few MeV for $m_u$. 

The $\eta^\prime$ mass and $m_u$ computations differ in that $m_{\eta^\prime}^2$ receives contributions from all topological charge sectors.  Thus there is no simple connection between the contributions except when $\rho \Lambda \ll 1$.

\bibliographystyle{utphys}
\bibliography{dinerefs}{}

\end{document}